\documentclass[sigconf,nonacm]{acmart}

\usepackage{booktabs}
\usepackage{graphicx}

\begin{document}

\title{From Agent Behaviour to Agent-Friendly Documentation}
\subtitle{An Empirical Study of How Coding Agents Discover, Read, and Write
Technical Documentation}

\author{Zhijun Gao}
\affiliation{%
  \institution{Peking University}
  \city{Beijing}
  \country{China}}
\email{gaozhijun@pku.edu.cn}

\author{Jing Chen}
\affiliation{%
  \institution{Peking University}
  \city{Beijing}
  \country{China}}
\email{2601210608@stu.pku.edu.cn}

\renewcommand{\shortauthors}{Gao and Chen}

\begin{abstract}
Technical documentation is written for human developers, but an increasing
share of software changes is now authored by autonomous coding agents. Which
documents these agents consult, when they consult them, and what follows remain
unknown. We conduct a behaviour-grounded study of
agents' interactions with documentation, combining two complementary public
datasets: 557 real agentic coding sessions from
SWE-chat~\cite{baumann_swechat_2026}, from which we extract 94{,}813 development
events, including 3{,}033 documentation interactions; and 33{,}097 agentic pull
requests from AIDev~\cite{li_aidev_2026}, for which we classify 690{,}260
file-level change records. Four findings challenge assumptions underlying
current documentation practice. First, agents' documentation work is dominated
by \emph{agent-facing} artefacts: agent instruction files and agent working notes
account for 60.5\% of all documentation interactions, whereas classical
technical documentation accounts for 10.6\% and API references for 1.3\%.
Second, the association between consultation and code editing remains
unresolved: the adjacent transition probability is 0.002 and the unadjusted
three-event lift is 1.05, whereas a stage-adjusted model places the association
above unity (OR 1.33 [1.09, 1.62]). Conversely, documentation creation is
elevated in the unadjusted analysis (lift 1.67), but its adjusted interval
includes unity. Third, no explicit documentation-based validation sequence was
observed, and consultation is associated with \emph{less} immediate testing
(lift 0.23, cluster CI 0.08--0.45; adjusted OR 0.39 [0.25, 0.60]). Fourth,
documentation consultation is self-initiated (70.2\%) far more often than it is
failure-driven (7.5\%), and documentation trails code rather than leading it:
among multi-commit pull requests that change both, code is touched first
4.7$\times$ more often than documentation. From these traces, we derive a
descriptive model of agent--documentation interaction that takes the form of a two-lobed
cycle rather than a linear journey, and we show that two widely assumed
properties of ``agent-friendly'' documentation --- actionability and
verifiability --- lack consistent behavioural support in this corpus. We
release our extraction pipeline, coding scheme, and event-level data.
\end{abstract}

\keywords{coding agents, software documentation, empirical software engineering,
agentic software development, trace analysis}

\maketitle

%===============================================================================
\section{Introduction}
\label{sec:intro}

Software documentation research has traditionally focused on a single audience:
the human developer. Documentation quality frameworks, staleness metrics, and
readability guidelines all presuppose a reader who forms intentions, gets
confused, and asks colleagues. That presupposition no longer always holds. A
substantial and growing fraction of code changes in open source is authored by
autonomous coding agents that read repositories, execute commands, and open pull
requests without a human in the loop at each step.

If agents are now a significant class of documentation consumers, documentation
design should be informed by what agents actually do. At present, such evidence
is lacking because their behaviour has not been measured. Emerging guidance on
``agent-friendly'' documentation --- write clear headings, provide runnable
examples, publish an \texttt{llms.txt} --- rests on intuition about how agents
\emph{ought} to behave rather than observation of how they do.

This paper takes the opposite route. Rather than proposing documentation
qualities and asking whether agents benefit, we observe agent behaviour in real
development sessions and derive documentation implications from these
observations. We ask three research questions:

\begin{itemize}
\item[\textbf{RQ1}] \emph{What do agents do with documentation?} Which document
types, at which stages of a task, through which behaviours?
\item[\textbf{RQ2}] \emph{What precedes and follows documentation interaction?}
Which events trigger consultation, and which development actions are more or
less likely after it?
\item[\textbf{RQ3}] \emph{Is the code--documentation loop bidirectional?} Do
agents consume and produce documentation, and in which order relative to code?
\end{itemize}

Our answers depart from the field's working assumptions in four ways, and the
departures are large rather than marginal.

\textbf{The dominant documentation genre is new.} We expected agents to consult
API references, architecture documents, and troubleshooting guides. Instead
60.5\% of observed documentation interaction targets artefacts that exist
\emph{because} of agents: instruction files such as \texttt{AGENTS.md} and
\texttt{CLAUDE.md} (35.4\%), and agent working notes --- plans,
\texttt{thoughts/} directories, brainstorms, verification logs (25.1\%). API
reference documentation, the focus of most documentation tooling, accounts for
1.3\% of interaction; troubleshooting documentation accounts for 0.4\%.

\textbf{The link from reading to coding is unresolved.} The commonly assumed
pattern --- \emph{read the docs, then write the code} --- is almost absent at the
adjacent-transition level:
$P(\text{edit code} \mid \text{read doc}) = 0.002$. At a three-event horizon,
the unadjusted lift is 1.05 over the session base rate, whereas the
stage-adjusted estimate is above unity (OR 1.33 [1.09, 1.62]). Documentation
reads are instead most often followed by reasoning (0.245) or further
documentation reads (0.270).

\textbf{Explicit documentation-based validation is not observed.} We observe no
instances in which documentation is explicitly used as an oracle against which
code is checked. Moreover, reading documentation is associated with \emph{less}
immediate testing (lift 0.23) and building (lift 0.15).

\textbf{Documentation is nearly as often an output as an input.} Production
(1{,}401 events) occurs at 0.87$\times$ the rate of consultation (1{,}615
events), and 41.5\% of agentic pull requests change documentation. However, the
direction is asymmetric in time: code precedes documentation 4.7$\times$ more
often than the reverse.

We contribute: (i) a behaviour-grounded characterisation of agent documentation
interaction across two datasets and two units of analysis; (ii) a released
extraction pipeline, evaluated against the dataset's own tool-call counts, that recovers
documentation events from four heterogeneous agent transcript formats (the
document-type classifier itself is \emph{not} human-validated); (iii) an
empirically derived interaction model that replaces the assumed linear journey
with a two-lobed cycle; and
(iv) documentation design implications, each tied to a specific measurement,
together with an explicit list of commonly asserted implications that our data
do \emph{not} support.

%===============================================================================
\section{Related Work}
\label{sec:related}

Five strands of literature converge in this paper, each resting on an assumption
that our data challenge: the documentation consulted by an agent or developer is
written for humans, concerns code, and was produced by someone else. The first
four strands expose this tension; the fifth establishes methodological precedent.

\subsection{Coding agents and LLM-based software-engineering agents}

The field's evaluation apparatus defines an agent's task as issue resolution.
SWE-bench scores an agent according to whether its patch makes tests pass
\cite{jimenez_swebench_2024}, and the architectures built to succeed on it are
described by their action spaces over files and shells \cite{yang_sweagent_2024,
wang_openhands_2024, zhang_autocoderover_2024}. Surveys map agent capabilities by
task type but do not include documentation as a category
\cite{liu_agents_survey_2024, jin_agents_se_2024}. Because no benchmark rewards
documentation work, evidence must come from observational data. Agentless
pipelines can match agentic ones without an exploratory reading loop
\cite{xia_agentless_2025}; our finding that reads lead predominantly to further
reading and reasoning (Section~\ref{sec:results}) therefore describes what
agents do, rather than what they necessarily need to do.

Two lines of observational work provide the closest precedents. Trajectory studies
compare successful and failed runs at the level of action sequences
\cite{code_agent_trajectories_2025, behavioral_drivers_2026}, motivating our
transition analysis, although neither codes documentation as an
artefact class. Artefact studies mine agent-authored pull
requests: adoption rates \cite{kuhrmann_agentic_adoption_2026}, activity over time
\cite{autonomous_contrib_wild_2026}, PR-description characteristics
\cite{prdesc_communicate_2026}, refactoring \cite{agentic_refactoring_2025},
logging \cite{agents_log_like_humans_2026}, and failure
causes \cite{failed_agentic_prs_2026}. The logging study provides a useful
template: it asks whether agents handle a non-functional concern in the same way
as humans. We ask the corresponding question about documentation and measure a
previously unreported quantity: agents produce documentation at 0.87 times the
rate at which they consult it (Section~\ref{sec:results}). Agent help-seeking has
been benchmarked as escalation to a human rather than consultation of an artefact
\cite{hilbench_ask_help_2026}; our distinction between self-initiated and
failure-driven interactions has no
prior baseline.

\subsection{Software documentation research before LLMs}

Research conducted before LLMs provides the clearest point of comparison.
Observed human developers rely more heavily on code and colleagues than on
documentation \cite{roehm_comprehend_2012, maalej_comprehension_2014,
ko_information_needs_2007}, consult selectively around staleness
\cite{lethbridge_documentation_2003}, frame needs as task-shaped questions
\cite{sillito_questions_2006, sillito_asking_2008}, and search the web for
external information \cite{sadowski_search_2015, xia_websearch_2017}.
Comprehension consumes most of developers' time \cite{xia_comprehension_2018}, without
documentation serving as its primary input. Against that baseline,
documentation interaction occurred in more than half of our sampled sessions
(56.7\%) and was predominantly self-initiated (70.2\%) --- the kind of behaviour
that documentation-engineering research has long sought to encourage among human
developers \cite{forward_relevance_2002, garousi_usage_2015,
zhi_cost_benefit_2015}.

The sharper contrast concerns \emph{which} documentation is consulted. This
literature centres on API references: what makes APIs hard to learn
\cite{robillard_apis_2009},
how API documentation fails \cite{uddin_apidoc_fails_2015}, how much exists on
the web \cite{parnin_measuring_api_2011}, how to navigate it by task
\cite{treude_navigate_2015}. Quality taxonomies are built from these artefact
types \cite{aghajani_issues_2019, aghajani_practitioners_2020}, whereas README
taxonomies focus on human-facing sections \cite{prana_readme_2018}. Our
distribution reverses this emphasis: API references account for 1.3\% of agents'
documentation interactions and troubleshooting documentation for 0.4\%, whereas
60.5\% involve agent-facing documents
(Section~\ref{sec:results}). The assumption that improving API reference quality
improves the consulting reader's outcomes may hold for humans, but API references
accounted for only 2.3\% of observable repository-local consultations in our
data. Research on onboarding barriers \cite{steinmacher_barriers_2015} and on
documentation creation \cite{dagenais_creating_2010} provide human comparison
points for our production-versus-consumption analysis.

\subsection{Code–documentation co-evolution}

The AIDev component builds on a mature co-change literature that documents
persistent problems:
comments are updated with code less often than they should be
\cite{fluri_coevolve_2007, fluri_analyzing_2009}, inconsistency is common and defect-associated
\cite{tan_icomment_2007, wen_inconsistency_2019, ibrahim_comment_update_2012}, comments break
silently under refactoring \cite{ratol_fragile_2017}, and documentation links decay
\cite{hata_links_2019}. We apply methods from the co-change mining tradition
\cite{zimmermann_mining_2005} to agentic PRs at the file level. Two
observations follow. First, the 41.5\% documentation-change rate is high relative
to rates reported in this literature for human comment maintenance. Second,
where order is observable, documentation trails code, reproducing the asymmetry
that automated
comment-update work presupposes \cite{panthaplackel_update_comments_2020}. When
this literature considers documentation being checked against code, the checking
is performed by a static analyser \cite{zhong_apidoc_errors_2013}; it does not
assume that an agent will perform the check, consistent with our observation of
no such events
(Section~\ref{sec:results}).

\subsection{LLMs, documentation, and agent context files}

Retrieval research treats documentation as model input: retrieval improves code
generation \cite{zhou_docprompting_2023, wang_coderagbench_2024}, as does
repository-level retrieval \cite{zhang_repocoder_2023, coderag_repo_2025}. This literature most directly
encodes the assumption that our findings challenge: documentation is valuable as
a retrievable API \emph{reference} injected on the model's behalf. The agents in
our data instead open instruction files they were told to follow and notes they
wrote themselves; these interactions are self-initiated rather than externally
retrieved.

The emerging 2025--2026 literature on context files addresses these artefacts,
but its conclusions remain unsettled. Descriptive studies characterise
\texttt{AGENTS.md} and \texttt{CLAUDE.md} as artefacts
\cite{agent_readmes_2025, manifests_claude_code_2025,
context_engineering_oss_2025}. Related work reports repository-content base rates
\cite{repo_contents_10k_2026} and examines plan artefacts
\cite{agent_plans_oss_2026}, the closest existing analogue to our working-notes
category. Evidence of effectiveness is mixed: reported efficiency gains
\cite{agentsmd_efficiency_2026}, mixed task-level results
\cite{evaluating_agentsmd_2026, context_files_ablation_2026}, and a finding that
random rules help as much as curated ones \cite{guardrails_beat_guidance_2026}.
Maintenance research reports staleness \cite{context_rot_2026} and documents
unbounded growth \cite{claudemd_growing_2026}; managing context through agent actions has emerged
as a technique \cite{context_as_tool_2025}. None of this work measures how often
agents consult these files relative to everything else they read. Our 35.4\%
instruction-file and 25.1\% working-notes figures provide that denominator, and
the conflicting effectiveness findings motivate reporting these figures
descriptively.

\subsection{Method precedent}

Process mining of software event logs is established
\cite{vanderaalst_manifesto_2012, poncin_process_mining_repos_2011}, as is
sequence mining of developers' IDE interaction streams
\cite{damevski_usage_smells_2017, kersten_mylyn_2006, cross_task_deps_2019}.
Human validation of our coding scheme, which we identify as the necessary next
step (Section~\ref{sec:threats}), would follow established reliability
apparatus --- Cohen's kappa \cite{cohen_kappa_1960} with conventional bands
\cite{landis_koch_1977}, or Krippendorff's alpha
\cite{krippendorff_reliability_2004}, reported following established guidance
\cite{mcdonald_reliability_2019, seaman_qualitative_1999}. We report no such
statistic here. Our primary intervals are cluster bootstraps; Wilson intervals
\cite{wilson_interval_1927} appear only as independence-assuming references; we
rely on their behaviour for small and zero counts \cite{brown_binomial_2001}.

%===============================================================================
\section{Study Design}
\label{sec:method}

\subsection{Datasets}

We use two public datasets with complementary units of analysis. Neither can
answer our questions alone: SWE-chat records development processes but not merge
outcomes, whereas AIDev records artefacts but retains no tool-use trajectories.

\textbf{SWE-chat}~\cite{baumann_swechat_2026} contains real agentic coding
sessions contributed by developers using several command-line agent tools. Each
session includes a complete transcript
with user messages, agent messages, tool calls, tool results, and code changes.
The release provides 5{,}850 sessions with transcripts (10.4\,GB); we sample 559
(Section~\ref{ssec:sampling}). This dataset is our source of \emph{process}
evidence: it answers RQ1 and RQ2 and provides within-session evidence for RQ3.

\textbf{AIDev}~\cite{li_aidev_2026} contains pull requests opened by coding
agents on public GitHub repositories, with commits, file-level diffs, reviews, and
timelines. The full corpus contains 932{,}791 PRs across 116{,}211 repositories;
we use the authors' curated subset of 33{,}596 PRs from 2{,}807 repositories with
more than 100 stars, together with its file-level commit-details table. All AIDev
figures below are \emph{derived} from that subset, so we report the filtering
steps explicitly.
The commit-details table yields 711{,}923 file$\times$commit rows after dropping
5{,}132 null-filename rows (0.72\%); excluding vendored paths removes 16{,}531
more (2.32\%), leaving \textbf{690{,}260} rows across 278{,}192 unique paths, of
which 29{,}597 are documentation paths. At the PR level, 33{,}097 of 33{,}596 PRs retain
at least one named non-vendored file (98.5\%); the 499 excluded are 16 with no
file-level rows, 475 with only null filenames, and 8 with only vendored files.
This dataset provides our \emph{artefact} evidence and answers RQ3 at scale.

The datasets describe related but distinct populations. We therefore use them as
complementary sources rather than as cross-validation and never pool their units
of analysis.

\subsection{Sampling}
\label{ssec:sampling}

From SWE-chat we draw a sample stratified by agent and session length across four
length strata (turn count $\leq 3$, 4--8, 9--18, $>18$), so that neither trivial
nor pathological sessions dominate. We oversample minority agents to permit
agent-specific estimation and exclude transcripts larger than 25\,MB (corpus maximum:
61.8\,MB). The sample is 559 sessions, 557 of which yielded parseable events.
Because allocation is non-proportional, agent-level statistics are reported
separately and are never pooled.

\subsection{Documentation identification}

We operationalise ``documentation'' using a two-tier classifier of
repository-relative file paths.

\emph{Tier 1} is deterministic. Filename and path rules assign one of 15
document types and two orthogonal flags. The \texttt{machine\_readable} flag
identifies OpenAPI, JSON Schema, and Protobuf artefacts, which serve as both API
documentation and executable specifications. The \texttt{vendored} flag
identifies third-party paths, which an
agent may read but the repository does not own. Non-documentation files receive a
\texttt{kind} in \{source, config, test, data, build, other\}, so a single function
supports both documentation identification and co-change analysis.

\emph{Tier 2} resolves ambiguous paths. Tier 1 placed 54\% of documentation
events in a residual category. Because this category was too large to leave
unresolved, we classified its 527 distinct paths with a language model: 500 were
labelled, covering 98.4\% of ambiguous events, and 27 were assigned using a
keyword fallback rule. This tier revealed \texttt{agent\_working\_note} as a
large, distinct category absent from our initial scheme.

Two decisions deserve emphasis. First, we do \emph{not} code interaction
\emph{purpose}, although it appeared in our initial scheme. Purpose cannot be
recovered from tool-call logs, and inferring intent from a file read would be
unfalsifiable; we report trigger, interaction type, and outcome instead. Second,
vendored paths are flagged rather than dropped: reading
\texttt{node\_modules/pkg/README.md} is a genuine documentation interaction even
though the repository does not own it.

\subsection{Event extraction}

SWE-chat uses four incompatible transcript formats, which we identified by
inspecting their structures.
Of the 559 sampled sessions, 406 use line-delimited JSON with content-block tool
calls, 100 use a single JSON document with a \texttt{parts} array, 43 use a
\texttt{\{type, payload\}} event log, and 10 use a \texttt{messages} array. Two
sessions yielded no events and were excluded, leaving an analytic sample of 557.
A separate extractor for each format emits a common schema based on a 20-symbol
alphabet spanning documentation and non-documentation actions, so documentation
events remain embedded in their original trajectory context.

Two extraction details materially affect the results and can bias this class of
study if left unreported. First, agents that route file operations through the
shell express edits as \texttt{apply\_patch} here-documents whose target paths
appear only inside the command text; without parsing these paths, one agent
family would have registered
\emph{zero} documentation events. Second, tool output arrives as a string in some
formats and a list or dictionary in others. Supporting these additional types
recovered nine sessions and 4{,}358 events, including 77 documentation events.
Our extracted tool-event counts matched SWE-chat's own
\texttt{tool\_call\_count} exactly in five of six spot-checked sessions.

\subsection{Coding scheme and derived measures}

Each documentation event is coded along four dimensions: \emph{document type} (15
rule categories plus the two additions in Section~\ref{sec:results});
\emph{interaction type} (Discover, Search, Read, Edit, Create); \emph{trigger},
using a four-event lookback; and \emph{outcome}, using success and failure
signals in tool output. Development
\emph{stage} follows a trajectory heuristic: orientation before the first write,
implementation from the first write, verification after a passing test or build,
debugging after a failure signal, and delivery after the last write when
version-control activity dominates. Every event retains its evidence string, so
labels are auditable.

For RQ2 we report \emph{lift} over a base rate rather than raw conditional
probability, because an action that is common throughout a session will also
appear frequently after consultation. We compare the probability of an action
within three events of a documentation \emph{consultation} (Read, Search, or
Discover; $n=1{,}615$ anchors) with the corresponding probability estimated from
every \emph{non-anchor} event in the same sessions ($n=93{,}198$; this baseline
includes documentation edits, which are not consultation anchors).

\textbf{Observation scope.} Our instrument observes
\emph{repository-local, file-based} documentation interactions: tool calls whose
target resolves to a repository path, plus rare explicit documentation retrieval
calls. It does not observe API websites read through a
browser, knowledge already in the model's weights, or in-source docstrings.
Context files loaded by the runtime at session start are visible only when the
agent later reads or edits them explicitly, so instruction-file counts are lower
bounds on \emph{exposure}. All claims concern this observable slice.

\textbf{Operationalisation of the interaction-cycle stages.} The candidate
stages in Table~\ref{tab:journey} constitute a second, coarser coding of the same
events. They are \emph{not} mutually exclusive, so counts do not sum to 3{,}033. Each is
an observable pattern:

\begin{itemize}
\item \emph{Orient}: events in the orientation stage (before the first write).
\item \emph{Discover}: events classified as Search or Discover ($282+5$).
\item \emph{Retrieve}: Read (1{,}328) plus documentation-tool calls (16),
  which represent retrievals but not file reads --- hence 1{,}344.
\item \emph{Interpret}: reads or searches followed immediately by reasoning.
\item \emph{Revisit}: a read immediately followed by another read.
\item \emph{Apply}: reads followed by an edit or read of the documented artefact.
\item \emph{Recover}: failure episodes whose first recovery action is reading
  documentation (109).
\item \emph{Contribute/Update}: interaction type Edit or Create (1{,}401).
\item \emph{Validate}: reads followed by a test or build run.
\item \emph{Escalate}: reads followed by a request to the user --- distinct from
  the nine \emph{Ask user} recovery actions in Table~\ref{tab:recovery}, which
  follow a \emph{failure}.
\end{itemize}

\subsection{Statistical treatment}

Because the sample is deliberately non-proportional, we report headline
proportions under three weighting schemes --- pooled events, session-equal, and
sessions reweighted to the corpus agent distribution (Table~\ref{tab:weights}).
The third corrects agent-level oversampling only: the sample is stratified by
agent \emph{and} session length, and the population margin for the length strata
is not recoverable from the released index, so we present no joint-stratum
estimate.

Because events nest within sessions and pull requests within repositories,
observations are not independent Bernoulli trials --- the largest AIDev
repository alone contributes 8{,}911 pull requests. We therefore obtain our
primary uncertainty estimates using a \emph{cluster bootstrap} (2{,}000
resamples), resampling whole sessions for SWE-chat and whole repositories for
AIDev. We compute percentile intervals using fixed seeds (13 for lift, 11 for transitions, 7 for
proportions), and each resample recomputes the pooled proportion from summed
within-cluster counts. For the RQ2 action windows we additionally fit a logistic
GEE (exchangeable working correlation, clustered by session) adjusting for
development stage, within-session position, log session length, and agent
family, because consultation is not evenly distributed across trajectory phases;
we report the adjusted association alongside the unadjusted lift, not in place
of it. Table~\ref{tab:cluster} compares the cluster intervals with Wilson
intervals, which assume independence; clustering widens every interval, by up to 14$\times$ on
the AIDev side. Wilson intervals remain in the per-analysis tables as an
independence-assuming reference, but are not our primary estimate.

We apply no correction for the number of strata examined and therefore do not
interpret small differences between adjacent strata.

%===============================================================================
\section{Results}
\label{sec:results}

\begin{figure*}[t]
\centering
\includegraphics[width=\textwidth]{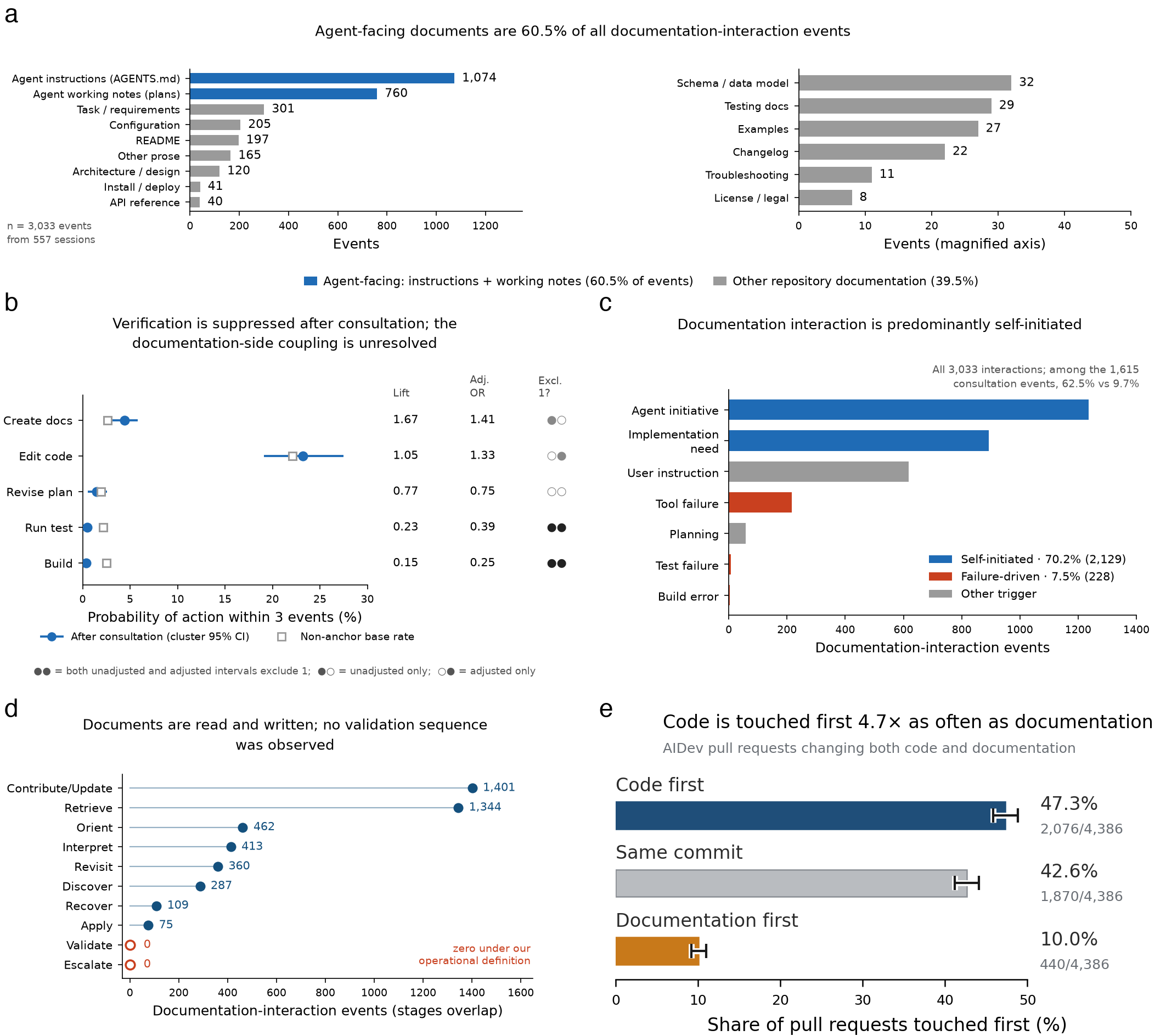}
\Description{Five panels summarising document types, post-consultation actions,
interaction triggers, candidate interaction stages, and code--documentation
ordering in pull requests.}
\caption{Agents' documentation interactions across 557 SWE-chat sessions
(panels a--d, $n=3{,}033$ documentation events) and 33{,}097 AIDev agentic pull
requests (panel e). (a) Agent-facing documents --- instruction files and agent
working notes --- account for 60.5\% of interactions, whereas API references
account for 1.3\%; the right-hand column uses a magnified axis. (b) Testing and building are less
frequent within three events of a consultation and remain so after adjustment;
the two authoring outcomes differ between the unadjusted lift and the
stage-adjusted odds ratio, so those associations remain unresolved. Intervals are
session-level cluster-bootstrap estimates. (c) Interaction is predominantly self-initiated (70.2\%
of all 3{,}033 interactions; 62.5\% of the 1{,}615 consultation events) rather
than failure-driven (7.5\%; 9.7\%); the axis counts all interactions. (d) Of ten
candidate stages, Validate and Escalate have zero events under our operational
definitions (open markers); stages overlap and do not sum to 3{,}033. (e) Among
pull requests that change both code and documentation, code is touched first
4.7$\times$ more often, although 42.6\% first touch both in a single commit, for
which no order is observable;
panel e uses a different
dataset and unit of analysis (pull requests), and its error bars are Wilson
intervals --- the repository cluster-bootstrap equivalents are in
Table~\ref{tab:cluster}.}
\label{fig:overview}
\end{figure*}

Our 557 sessions contain 94{,}813 events, of which 3{,}033 (3.2\%) are
documentation interactions. Such interactions are common but not universal: 316
sessions (56.7\%, cluster 95\% CI 52.6--60.5\%) contain at least one.
Figure~\ref{fig:overview} summarises the principal results.

\subsection{RQ1: Documentation Types, Task Stages, and Interaction Behaviours}

\subsubsection{Task stages}

Documentation interaction is distributed across the whole trajectory rather than
concentrated at task start. By development stage, 54.4\% of documentation events
occur during debugging, 27.2\% during implementation, 15.2\% during orientation,
3.0\% during verification, and 0.1\% at delivery.

The stage heuristic is sticky: once a failure signal appears, the session remains
in debugging until a test or build passes, inflating that share. We therefore
advance only the robust negative claim: documentation consultation is \emph{not}
confined to the orientation phase. A model that treats documentation solely as a
task-start activity is therefore inconsistent with our observations. The
distribution across stages should not be interpreted as a precise allocation.

\subsubsection{Document types}

Table~\ref{tab:doctypes} presents the full distribution, which is the paper's
central empirical result. Two categories of \emph{agent-facing} documents dominate.
Agent instruction files --- including \texttt{AGENTS.md}, \texttt{CLAUDE.md},
\texttt{SKILL.md}, and rule files for Cursor and Copilot --- account for 1{,}074
events (35.4\%). Agent working notes
--- plans, \texttt{thoughts/} directories, brainstorms, and review logs that the
agent writes for its own use --- account for 760 (25.1\%). Together, these two
categories account for 1{,}834 of 3{,}033 events, or 60.5\%.

Our primary contrast is therefore between agent-facing artefacts (60.5\% of
events, session-cluster 95\% CI 53.9--66.5\%; 55.1\% under agent reweighting)
and all other repository documentation (39.5\%; 44.9\% under agent reweighting).
As a secondary contrast,
the nine genres at the traditional core of documentation research account for
323 events (10.6\%). This boundary is contestable: adding README, configuration,
and requirements documentation raises the share to 33.8\%. We therefore make no
claim beyond the extreme categories: API references account for 40 events
(1.3\%), and troubleshooting documentation --- expected to be the debugging
resource --- accounts for 11 (0.4\%).

The \texttt{agent\_working\_note} category did not exist in our initial scheme.
It emerged from Tier-2 classification of unresolved paths, which were dominated
by agent-authored planning and reasoning documents. Whether these files count as
technical documentation is a definitional choice that we make explicit: we
include them as durable prose artefacts about the software that are committed to
the repository and readable by the next actor, but we always report
them separately so readers drawing the boundary differently can recompute every
share (excluding them, agent instruction files alone are 1{,}074 of 2{,}273
events, 47.2\%).

\begin{table}[t]
\centering
\small
\caption{Documentation-interaction events by document type ($n=3{,}033$ events
across 557 sessions). Agent-facing categories are marked $\dagger$. Together,
the two agent-facing categories account for 60.5\% of interactions; the nine
categories constituting classical technical documentation (API reference,
troubleshooting, architecture, schema, installation, examples, testing,
contributing, and changelog) account for 10.6\%.}
\label{tab:doctypes}
\begin{tabular}{lrr}
\toprule
Document type & Events & Share \\
\midrule
Agent instructions$^\dagger$ & 1,074 & 35.4\% \\
Agent working notes$^\dagger$ & 760 & 25.1\% \\
Task / requirements & 301 & 9.9\% \\
Configuration & 205 & 6.8\% \\
README & 197 & 6.5\% \\
Other prose (residual) & 165 & 5.4\% \\
Architecture / ADR & 120 & 4.0\% \\
Install / deploy & 41 & 1.4\% \\
API reference & 40 & 1.3\% \\
Schema & 32 & 1.1\% \\
Testing docs & 29 & 1.0\% \\
Examples & 27 & 0.9\% \\
Changelog & 22 & 0.7\% \\
Troubleshooting & 11 & 0.4\% \\
License / legal & 8 & 0.3\% \\
Contributing & 1 & 0.0\% \\
\midrule
\textit{Agent-facing subtotal}$^\dagger$ & \textbf{1,834} & \textbf{60.5\%} \\
\textit{Classical technical documentation} & 323 & 10.6\% \\
\bottomrule
\end{tabular}
\end{table}

Because working notes might be mostly written whereas API references are mostly
read, Table~\ref{tab:readwrite} splits each document type into consultation
(Read, Search, Discover) and production (Edit, Create). Agent-facing dominance
holds on both sides: 57.4\% of consultation and 63.7\% of production, compared
with 2.3\% of consultation for API references. Configuration files are the most
asymmetric type (195 consultations, 10 productions).

\subsubsection{Interaction behaviours}

We observe five interaction types: Read (1{,}328 events), Edit (1{,}007),
Create (394), Search (282), and Discover (5). Of the remaining 17 events, 12 are
failed tool calls and 5 do not fit any interaction type. These events carry
document-type and trigger labels and contribute to those distributions but not
to interaction-type statistics. Production (Edit + Create = 1{,}401)
occurs at 0.87$\times$ the rate of consultation (Read + Search + Discover =
1{,}615).

Three interaction types from our initial scheme are \emph{not} attested: Compare
(reading two documents against each other), Follow-reference (navigating a link
between documents), and Verify (checking code against documentation). These may
occur inside model reasoning, which our instrument cannot see, but they do not
appear as tool-call behaviour, so we remove them rather than report them as
rare.

The most common recurrent transitions are as follows (session cluster-bootstrap
95\% CIs; 2{,}000 resamples).

\begin{align*}
P(\text{read doc} \mid \text{read doc}) &= 0.270 && [0.232, 0.307],\\
P(\text{reasoning} \mid \text{read doc}) &= 0.245 && [0.205, 0.295],\\
P(\text{edit doc} \mid \text{read doc}) &= 0.107 && [0.081, 0.132],\\
P(\text{edit doc} \mid \text{edit doc}) &= 0.350 && [0.288, 0.404].
\end{align*}
The first estimate indicates that documentation reads occur in \emph{runs}
rather than in isolation.

At the adjacent-transition level, the commonly assumed pattern --- read
documentation, then write code --- is nearly absent:
\[
P(\text{edit code} \mid \text{read doc}) = 0.002
\quad [0.000, 0.005].
\]
This estimate represents three occurrences among 1{,}328 documentation reads. A read is much more likely
to be followed by reasoning or further reading than by an immediate code edit.

\subsection{RQ2: Antecedents and Subsequent Actions}

\subsubsection{Triggers}

Table~\ref{tab:triggers} presents the trigger distribution for all 3{,}033
documentation interactions. In aggregate, 2{,}129 events (70.2\%, session-cluster
95\% CI 66.7--73.3\%) are
self-initiated --- the agent's own initiative (1{,}236) or an implementation need
from its current work (893). Failure-driven interaction accounts for 228 events
(7.5\%, CI 6.0--9.3\%): tool failure (217), test failure (7), build error (4).
User instructions account for 618 events (20.4\%). When the analysis is restricted
to consultation ($n=1{,}615$), the pattern persists: 62.5\% are self-initiated
(1{,}010) versus 9.7\%
failure-driven (156).

Self-initiated interactions outnumber failure-driven interactions by
9.3$\times$ (6.5$\times$ for consultation alone). Thus, rather than consulting
documentation primarily when something goes wrong, agents consult it during
routine task progress.

\begin{table}[t]
\centering
\small
\caption{RQ2: what triggers documentation \emph{interaction} ($n=3{,}033$ interactions; consultation events alone are $n=1{,}615$, for which the self-initiated share is 62.5\%). Self-initiated triggers outnumber failure-driven ones by 9.3$\times$.}
\label{tab:triggers}
\begin{tabular}{lrr}
\toprule
Trigger & Events & Share \\
\midrule
Agent initiative & 1,236 & 40.8\% \\
Implementation need & 893 & 29.4\% \\
User instruction & 618 & 20.4\% \\
Tool failure & 217 & 7.2\% \\
Planning & 58 & 1.9\% \\
Test failure & 7 & 0.2\% \\
Build error & 4 & 0.1\% \\
\midrule
\textit{Self-initiated} & \textbf{2,129} & \textbf{70.2\%} \\
\textit{Externally instructed} & 618 & 20.4\% \\
\textit{Failure-driven} & 228 & 7.5\% \\
\bottomrule
\end{tabular}
\end{table}

\subsubsection{Actions following consultation}

Table~\ref{tab:lift} reports the probability of each action within three events
of a documentation \emph{consultation} (Read, Search, or Discover; $n=1{,}615$
anchors), compared with the rate at non-anchor events in the same sessions
($n=93{,}198$).

Two verification actions are less frequent within the next three events and both
survive adjustment: running a test (lift 0.23, cluster CI 0.08--0.45; adjusted OR
0.39 [0.25, 0.60]) and building (0.15, CI 0.02--0.33; OR 0.25 [0.14, 0.44]). The
unadjusted and adjusted estimates differ for both authoring outcomes.
Documentation creation is elevated in the unadjusted analysis (lift
1.67, CI 1.14--2.31), but its adjusted interval includes unity (OR 1.41 [0.98,
2.02]). Conversely, code editing is indistinguishable from the baseline in the
unadjusted analysis (lift 1.05, CI 0.86--1.27), yet elevated after controlling
for stage (OR 1.33 [1.09, 1.62]).
Because consultation concentrates in particular trajectory phases, this
discrepancy is consistent with the stage confounding flagged as a threat. We
therefore treat the lower frequency of test and build activity as the finding,
and any consultation-to-authoring or
consultation-to-code coupling as unresolved by these data.

These results provide no consistent support for the simple mechanism in which
documentation consultation is followed by implementation and verification. The
code-editing association depends on adjustment, and test and build activity are
lower after consultation. No positive downstream-action association is robust
across both analyses.

\begin{table*}[t]
\centering
\small
\setlength{\tabcolsep}{3.5pt}
\caption{Actions within three events of a documentation consultation ($n_c=1{,}615$
anchors) compared with the non-anchor base rate ($n_b=93{,}198$), with session
cluster-bootstrap intervals (2{,}000 resamples, percentile method) for the risk
difference $\Delta p$ and the lift ratio, and an adjusted association from a
logistic GEE with exchangeable working correlation clustered by session and
adjusted for development stage, within-session position, session length, and
agent family. Unadjusted and adjusted estimates differ for the two authoring
outcomes: documentation creation is elevated before adjustment, but its adjusted
interval includes 1; code editing is indistinguishable from the baseline before
adjustment, but is elevated after controlling for stage. Only the lower
frequencies of testing and building are robust to both analyses. We therefore
treat these reductions as the finding and the authoring associations as
unresolved.}
\label{tab:lift}
\begin{tabular}{lrrrrr}
\toprule
Action & After consult. & Base & $\Delta p$ [95\% CI] & Lift [95\% CI] & Adj.\ OR [95\% CI] \\
\midrule
Create documentation & 0.044 & 0.026 & +0.0176 [+0.0040, +0.0320] & 1.67 [1.14, 2.31] & 1.41 [0.98, 2.02] \\
Edit code & 0.232 & 0.221 & +0.0113 [-0.0325, +0.0587] & 1.05 [0.86, 1.27] & 1.33 [1.09, 1.62] \\
Revise plan & 0.015 & 0.019 & -0.0044 [-0.0124, +0.0042] & 0.77 [0.35, 1.24] & 0.75 [0.43, 1.30] \\
Run test & 0.005 & 0.022 & -0.0169 [-0.0250, -0.0094] & 0.23 [0.08, 0.45] & 0.39 [0.25, 0.60] \\
Build & 0.004 & 0.025 & -0.0214 [-0.0292, -0.0143] & 0.15 [0.02, 0.33] & 0.25 [0.14, 0.44] \\
\bottomrule
\end{tabular}
\end{table*}

\subsubsection{Failure recovery}

We identify 2{,}034 failure episodes (a failing test, a failed build, or a tool
error) and
classify the agent's first subsequent action. Table~\ref{tab:recovery} reports
frequency and resolution rate.

Reading documentation is the first recovery move in 109 of 2{,}034 episodes
(5.4\%). Agents far more often read code (631), retry the same action
(404), take no recovery action within the horizon (318), or edit directly (312).
$P(\text{read doc} \mid \text{tool error}) = 0.020$.

Documentation-based recovery has the highest point estimate of the resolution rate
(7/11 = 63.6\%), but its 95\% interval is 35.4--84.8\% and overlaps every other
strategy. We therefore report this as suggestive and explicitly \emph{not} a
finding: 11 episodes with an observable outcome cannot support a ranking of
recovery strategies. Establishing whether documentation-based recovery is
genuinely more effective is therefore an important motivation for a larger sample
(Section~\ref{sec:threats}).

\begin{table}[t]
\centering
\small
\caption{Failure recovery: the first action after each of 2{,}034 failure
episodes and the fraction of episodes subsequently resolved. Resolution can be
computed only for episodes followed by an observable test or build outcome, so
the resolution denominators are smaller than the frequency counts.
Documentation-based recovery has the highest point estimate but the widest
interval; it overlaps every other strategy, so we draw no conclusion about the
relative ranking.}
\label{tab:recovery}
\footnotesize
\setlength{\tabcolsep}{4pt}
\begin{tabular}{lrrc}
\toprule
First recovery action & $n$ (\%) & Resolved & 95\% CI \\
\midrule
Read code & 631 (31.0) & 39/159 & [.18, .32] \\
Retry directly & 404 (19.9) & 48/192 & [.19, .32] \\
No action within horizon & 318 (15.6) & 5/67 & [.03, .16] \\
Edit directly & 312 (15.3) & 45/160 & [.22, .36] \\
Search code & 251 (12.3) & 17/73 & [.15, .34] \\
\textbf{Read documentation} & 109 (5.4) & 7/11 & [.35, .85] \\
Ask user & 9 (0.4) & --- & --- \\
\bottomrule
\end{tabular}
\end{table}

\subsection{RQ3: Documentation Consumption, Production, and Code--Document Order}

Together, the two datasets show that agents consume and produce documentation.

\textbf{Within sessions.} Of the 316 sessions with documentation activity, 184
(58.2\%) both read and wrote; 102 (32.3\%) only read; 28 (8.9\%) only wrote; and 2
(0.6\%) contained only unclassified events, so no read/write status could be assigned.

\textbf{Across pull requests.} Table~\ref{tab:aidev} presents the artefact-level
results for the curated AIDev subset~\cite{li_aidev_2026}. Of the 33{,}097
pull requests, 13{,}750 change documentation (41.5\%, repository-cluster 95\% CI
35.8--45.4\%). Code--documentation co-change occurs in 32.0\% (CI
24.4--38.9\%) of all PRs, which is
37.0\% of the 28{,}574 code-touching PRs; a further 9.6\% (CI 6.1--14.5\%) change
documentation only. These PRs are heavily clustered --- the ten largest
repositories account for 44.7\% --- so the intervals are roughly an order of magnitude
wider than the Wilson equivalents (Table~\ref{tab:cluster}).

\textbf{Direction.} Among the 4{,}386 multi-commit pull requests where ordering is
observable, code is touched first in 47.3\%, both are first touched in the same
commit in 42.6\%, and documentation is touched first in 10.0\%. When the two change in different
commits, code comes first in 82.5\% of orderable cases (2{,}076/2{,}516;
repository-cluster CI 78.7--86.0\%): documentation follows code far more often
than it leads.

\textbf{Merge rates are not distinguishable.} Among closed PRs, 81.1\% of those
touching documentation were merged, compared with 75.0\% of code-only PRs. Under
repository clustering, the intervals overlap substantially (71.3--85.6\% and
64.9--81.1\%), so we draw no conclusion; assuming independence would make the
difference appear more decisive than the clustered analysis supports.

\textbf{Agents edit their own instructions.} Among the most-changed individual
documentation files in AIDev are \texttt{AGENTS.md} (692 PRs),
\texttt{CLAUDE.md} (362), and \texttt{copilot-instructions.md} (287). Agents
modify files that shape agent behaviour, closing a second loop --- from agent
output back to agent input --- that existing documentation models do not capture.

\begin{table}[t]
\centering
\small
\caption{Artefact-level results for RQ3 (AIDev pull requests): (a)
documentation involvement among all analysable agentic PRs; (b) the artefact
touched first, restricted to multi-commit PRs that change both artefact types and
for which order is observable; and (c) merge rate by documentation involvement,
reported as an association rather than a causal effect. All intervals are Wilson
95\% CIs.}
\label{tab:aidev}
\footnotesize
\setlength{\tabcolsep}{4pt}
\begin{tabular}{lrc}
\toprule
& $k/n$ & Prop.\ [95\% CI] \\
\midrule
\multicolumn{3}{l}{\textit{(a) Documentation involvement}} \\
\quad Code only & 17,988/33,097 & .543 [.538, .549] \\
\quad Code + documentation & 10,586/33,097 & .320 [.315, .325] \\
\quad Documentation only & 3,164/33,097 & .096 [.092, .099] \\
\quad Neither (build/config/data) & 1,359/33,097 & .041 [.039, .043] \\
\midrule
\multicolumn{3}{l}{\textit{(b) First artefact touched}} \\
\quad Code first & 2,076/4,386 & .473 [.459, .488] \\
\quad Same commit & 1,870/4,386 & .426 [.412, .441] \\
\quad Documentation first & 440/4,386 & .100 [.092, .110] \\
\midrule
\multicolumn{3}{l}{\textit{(c) Merge rate}} \\
\quad Touches documentation & 10,303/12,707 & .811 [.804, .818] \\
\quad Code only & 12,614/16,815 & .750 [.744, .757] \\
\bottomrule
\end{tabular}
\end{table}

\begin{table*}[t]
\centering
\small
\setlength{\tabcolsep}{3pt}
\caption{Cluster-bootstrap intervals (2{,}000 resamples) for the headline
proportions, resampling \emph{sessions} for SWE-chat statistics and
\emph{repositories} for AIDev statistics, alongside Wilson intervals that
assume independent trials. The ratio column gives cluster-interval width divided
by Wilson-interval width. Clustering widens every interval, especially for AIDev, where the largest
repository contributes 8{,}911 pull requests and the ten largest contribute
44.7\%. We report cluster intervals as the primary uncertainty estimates.}
\label{tab:cluster}
\begin{tabular}{lrcccr}
\toprule
Statistic & $k/n$ & Prop. & Cluster-bootstrap 95\% CI & Wilson 95\% CI & Width ratio \\
\midrule
\multicolumn{6}{l}{\textit{Sessions as clusters (SWE-chat)}} \\
\quad Agent-facing, all interactions & 1,834/3,033 & 0.605 & [0.539, 0.665] & [0.587, 0.622] & 3.6 \\
\quad \quad consultation events & 927/1,615 & 0.574 & [0.508, 0.634] & [0.550, 0.598] & 2.6 \\
\quad \quad production events & 893/1,401 & 0.637 & [0.553, 0.713] & [0.612, 0.662] & 3.2 \\
\quad Self-initiated, all interactions & 2,129/3,033 & 0.702 & [0.667, 0.733] & [0.685, 0.718] & 2.0 \\
\quad Failure-driven, all interactions & 228/3,033 & 0.075 & [0.060, 0.093] & [0.066, 0.085] & 1.8 \\
\quad Sessions with doc event & 316/557 & 0.567 & [0.526, 0.605] & [0.526, 0.608] & 1.0 \\
\quad Doc-first recovery & 109/2,034 & 0.054 & [0.038, 0.073] & [0.045, 0.064] & 1.8 \\
\multicolumn{6}{l}{\textit{Repositories as clusters (AIDev)}} \\
\quad Documentation changed (PRs) & 13,750/33,097 & 0.415 & [0.358, 0.454] & [0.410, 0.421] & 9.1 \\
\quad Code--doc co-change (PRs) & 10,586/33,097 & 0.320 & [0.243, 0.389] & [0.315, 0.325] & 14.4 \\
\quad Code first, different commits & 2,076/2,516 & 0.825 & [0.787, 0.860] & [0.810, 0.839] & 2.5 \\
\quad Merge rate, doc-touching & 10,303/12,707 & 0.811 & [0.713, 0.856] & [0.804, 0.818] & 10.5 \\
\quad Merge rate, code-only & 12,614/16,815 & 0.750 & [0.649, 0.811] & [0.744, 0.757] & 12.4 \\
\bottomrule
\end{tabular}
\end{table*}

\subsection{Robustness and Variation}

Because we oversample minority agents (Section~\ref{sec:method}), pooled
statistics overweight long sessions and minority agents relative to their corpus
prevalence. Table~\ref{tab:weights} reports each headline
proportion under pooled-event, session-equal, and agent-reweighted estimation.
The agent-facing share of all interaction moves from 60.5\% to 54.7\% and
55.1\%; self-initiated from 70.2\% to 63.2\% and 61.9\%. Split by side,
agent-facing \emph{consultation} moves from 57.4\% to 50.5\% and 50.1\% --- at
the 50\% boundary, so under either correction agent-facing documents account for
about half of consultation rather than a clear majority --- while
\emph{production} rises from 63.7\% to 67.1\% and 66.3\%. The broad pattern is
stable, but whether agent-facing documents constitute a majority of consultation
events depends on the weighting. The weighting corrects only the agent margin,
not the joint agent-by-length strata
(Section~\ref{sec:threats}).

\begin{table}[t]
\centering
\footnotesize
\setlength{\tabcolsep}{3.5pt}
\caption{Consultation (Read, Search, Discover; $n_c=1{,}615$) versus production (Edit, Create; $n_p=1{,}401$) events by document type. Agent-facing categories are marked $\dagger$; they account for 57.4\% of consultation and 63.7\% of production. Configuration files are consulted far more than they are produced; agent working notes are consulted and produced in nearly equal measure. ``Other low-frequency'' pools troubleshooting, license/legal, and contributing, so the columns sum to $n_c$ and $n_p$.}
\label{tab:readwrite}
\begin{tabular}{lrrrr}
\toprule
Document type & Cons. & \% & Prod. & \% \\
\midrule
Agent instructions$^\dagger$ & 545 & 33.7 & 526 & 37.5 \\
Agent working notes$^\dagger$ & 382 & 23.7 & 367 & 26.2 \\
Task / requirements & 129 & 8.0 & 170 & 12.1 \\
Configuration & 195 & 12.1 & 10 & 0.7 \\
README & 89 & 5.5 & 107 & 7.6 \\
Other prose & 72 & 4.5 & 93 & 6.6 \\
Architecture / ADR & 69 & 4.3 & 51 & 3.6 \\
Install / deploy & 24 & 1.5 & 17 & 1.2 \\
API reference & 37 & 2.3 & 3 & 0.2 \\
Schema & 16 & 1.0 & 16 & 1.1 \\
Testing docs & 19 & 1.2 & 10 & 0.7 \\
Examples & 17 & 1.1 & 10 & 0.7 \\
Changelog & 11 & 0.7 & 11 & 0.8 \\
Other low-frequency & 10 & 0.6 & 10 & 0.7 \\
\midrule
\textit{Total} & 1,615 & 100.0 & 1,401 & 100.0 \\
\textit{Agent-facing}$^\dagger$ & \textbf{927} & \textbf{57.4} & \textbf{893} & \textbf{63.7} \\
\bottomrule
\end{tabular}
\end{table}

\begin{table}[t]
\centering
\footnotesize
\setlength{\tabcolsep}{3pt}
\caption{Sensitivity of headline proportions to weighting. The event-weighted
estimate pools all events (long sessions weigh more); the session-equal estimate
averages per-session shares; and the agent-weighted estimate reweights sessions
to the corpus agent distribution (83.8\%
Claude Code, 4{,}852/5{,}790 labelled sessions; 61 of the 5{,}851 released
sessions are excluded from the reweighting --- 52 carry no agent label and 9
belong to agents with fewer than 20 sessions),
correcting our deliberate oversampling of minority agents. Because the sample was also
stratified by session length, this corrects agent-level oversampling only and is not a
full joint-stratum estimate. All three statistics describe the stratified sample;
the broad pattern is stable across all three weightings.}
\label{tab:weights}
\begin{tabular}{lccc}
\toprule
Statistic & Event & Sess.-eq. & Agent-wt. \\
\midrule
Agent-facing, all interactions & 60.5\% & 54.7\% & 55.1\% \\
\quad consultation events only & 57.4\% & 50.5\% & 50.1\% \\
\quad production events only & 63.7\% & 67.1\% & 66.3\% \\
Self-initiated, all interactions & 70.2\% & 63.2\% & 61.9\% \\
Sessions w/ doc event & --- & 56.7\% & 59.8\% \\
\bottomrule
\end{tabular}
\end{table}

Restricting the RQ2 anchors to reads alone ($n=1{,}328$) does not change any
conclusion: code-edit lift increases from 1.05 to 1.10, documentation-creation
lift from 1.67 to 1.83, test-running lift from 0.23 to 0.24, and build lift from
0.15 to 0.18.

\subsubsection{Variation across agents}

Table~\ref{tab:agents} reports session-level documentation rates per agent. The
rates differ substantially, from 62.6\% (Claude Code, 238/380) to 37.2\%
(Codex, 16/43); one agent has a rate of 0/11.

We caution against interpreting these as behavioural differences. One agent routes
nearly all file work through shell commands, so its interactions are visible only
if the extractor parses paths from command text. Before these paths were parsed,
the agent registered zero events. Cross-agent comparison is confounded with extraction
coverage.

\begin{table}[t]
\centering
\footnotesize
\setlength{\tabcolsep}{4pt}
\caption{Fraction of sessions containing at least one documentation interaction,
by agent. Sampling was deliberately non-proportional, so these per-agent
estimates must not be pooled. The zero estimate for one agent is based on 11
sessions and does not constitute evidence of absence.}
\label{tab:agents}
\begin{tabular}{lrrc}
\toprule
Agent & With docs & Sessions & Rate [95\% CI] \\
\midrule
Gemini CLI & 9 & 12 & .750 [.468, .911] \\
Agent & 8 & 12 & .667 [.391, .862] \\
Claude Code & 238 & 380 & .626 [.577, .673] \\
OpenCode & 45 & 99 & .455 [.360, .552] \\
Codex & 16 & 43 & .372 [.244, .521] \\
Cursor & 0 & 11 & .000 [.000, .259] \\
\bottomrule
\end{tabular}
\end{table}

%===============================================================================
\section{A Trace-Derived Descriptive Model of Agent Documentation Interaction}
\label{sec:model}

Our study began with a hypothesised linear journey, adapted from accounts of
human developer information seeking:
\begin{center}
Discover $\rightarrow$ Retrieve $\rightarrow$ Interpret $\rightarrow$\\
Apply $\rightarrow$ Validate $\rightarrow$ Update
\end{center}
Table~\ref{tab:journey} summarises the evidence for each candidate stage. The
data do not support the linear model for three reasons.

\begin{table}[t]
\centering
\small
\caption{Evidence for each candidate stage of the documentation-interaction
cycle ($n=3{,}033$ documentation events). Two stages in the initial scheme are
entirely unattested.}
\label{tab:journey}
\begin{tabular}{lrl}
\toprule
Candidate stage & Events & Status \\
\midrule
Contribute/Update & 1,401 & strongly attested \\
Retrieve & 1,344 & strongly attested \\
Orient & 462 & attested \\
Interpret & 413 & attested \\
Revisit & 360 & attested \\
Discover & 287 & attested \\
Recover & 109 & attested, weak \\
Apply & 75 & weak \\
Validate & 0 & \textbf{not attested} \\
Escalate & 0 & \textbf{not attested} \\
\bottomrule
\end{tabular}
\end{table}

\textbf{Two stages are unattested.} No events match Validate or Escalate under
our operational definitions: consultation followed by a test or build run, or by
a request to the user. These are zeros for the defined \emph{patterns}, not
evidence that no validation of any form occurs
(Section~\ref{sec:threats}).

\textbf{Apply is weakly attested.} Only 75 events show documentation reading
followed by action on the documented artefact, and the lift analysis
(Table~\ref{tab:lift}) estimates an unadjusted lift of 1.05 for subsequent code
editing, although the adjusted odds ratio is above unity. The link that the
linear model treats as its central step is therefore the one our data leave
least settled.

\textbf{The terminal stage is the largest.} Contribute/Update, which appears
last in the linear model, is the single largest category, with 1{,}401 events ---
more than Retrieve (1{,}344).

\subsection{A trace-derived descriptive model: the two-lobed cycle}

The transition structure shows two loosely coupled activity lobes.

The \textbf{consultation lobe} (Orient $\rightarrow$ Discover $\rightarrow$
Retrieve $\rightarrow$ Interpret) is internally recurrent: its strongest
transition is Retrieve back to itself at 0.270 (CI 0.232--0.307) and its
strongest outgoing transition is to reasoning at 0.245 (CI 0.205--0.295). Agents
therefore circulate within this lobe and move into reasoning more often than into
immediate action.

The \textbf{production lobe} (Contribute/Update) contains more documentation
events than any other stage. The unadjusted analysis suggests an association from
consultation into this lobe: documentation creation is elevated after
consultation (lift 1.67). Its stage-adjusted interval, however, includes unity
(Table~\ref{tab:lift}).

Neither connection is consistent across specifications. The unadjusted estimates
are 1.67 for consultation $\rightarrow$ documentation and 1.05 for consultation
$\rightarrow$ code; the corresponding adjusted ORs are 1.41 and 1.33. The
documentation interval excludes unity only before adjustment, whereas the code
interval excludes unity only after adjustment. Failure feeds into the
consultation lobe only rarely (5.4\% of failure episodes), and no observed edge
runs from either lobe into validation in the recorded tool-call traces.

Our revised account is therefore not a pipeline from an information need to
validated implementation. Instead, agents' interaction with documentation is a
recurrent consultation process that produces reasoning and further
documentation and is only loosely coupled to a largely independent
code-modification process.

\subsection{Why the difference from human developers matters}

The human information-seeking literature often describes developers who consult
documentation, apply what they learn, and check the result --- the loop our
linear model encoded. In our traces, agents perform the first step, but the
checking step is not observed, and documentation \emph{authorship} follows
consultation at a measurable rate.

Two mechanisms plausibly explain this; distinguishing between them requires
future work.
Agents may externalise reasoning to files because their context windows are
bounded, making documentation a form of working memory rather than a reference;
the prominence of plans and \texttt{thoughts/} directories is consistent with
this possibility. Alternatively, they may not
validate against prose because a cheaper oracle, the test suite, is invoked
directly. In either case, we did not observe prose functioning as a specification.

%===============================================================================
\section{Implications for Documentation Design}
\label{sec:implications}

We state implications only where a specific measurement supports them, and we
separately list the implications commonly asserted in this area that our data
do \emph{not} support. The second list matters as much as the first: this study
provides initial behavioural evidence on several of these questions but does not
support some widely repeated advice.

\subsection{Supported implications}

\textbf{Agent instruction files are the most frequently used documentation surface.}
They are the most frequently used document type (1{,}074 events, 35.4\%)
and are among the most frequently changed files in agentic pull requests. API
references, by contrast, receive 40 events (1.3\%). Instruction files receive
roughly 27$\times$ as many interactions as API references. For projects
allocating finite documentation resources to support agentic contributors, this
difference suggests prioritising the correctness and clarity of instruction files.

\textbf{Local retrievability warrants particular attention.} Documentation reads
are frequently followed by further reads (transition probability 0.270), whereas
Follow-reference is entirely unattested. This pattern motivates studying
self-contained documents with locally retrievable structure, rather than
assuming that agents navigate richly cross-linked documentation. It does not,
however, establish that link hygiene has no behavioural consequences.

\textbf{Agent-authored documents create a new maintenance surface.}
Documentation creation has an unadjusted lift of 1.67 after consultation,
although the stage-adjusted interval includes unity. Agent working notes account
for 25.1\% of all documentation interactions. We measured their volume and
modification, not their maintenance cost, staleness, or inconsistency. Plans,
\texttt{thoughts/} directories, and verification logs accumulate in repositories
as durable artefacts. Repository hygiene tooling, code review checklists, and
documentation quality metrics currently have no category for them.

\textbf{Executable documentation offers a testable route to specification.}
No explicit documentation-based validation sequence was observed, and
consultation is associated with less immediate testing (lift 0.23, cluster CI
0.08--0.45). Making such a check observable plausibly requires artefacts an agent
can execute --- runnable examples, doctests, schema contracts --- rather than
prose that an agent must be trusted to honour. This proposal is a hypothesis for
intervention studies, not a finding of the present study.

\textbf{Documentation rarely appeared as the first recovery resource.} It was the
first recovery action in 109 of 2{,}034 failure episodes (5.4\%, cluster CI
3.8--7.3\%), and
troubleshooting documents specifically account for 11 events across the entire
corpus. Whatever value troubleshooting guides have for human developers, they
are not a prominent part of the agent recovery behaviour observed here.

\subsection{Implications our data do \emph{not} support}

\textbf{Actionability.} The claim that documentation should be written so agents
can act on it directly presumes a read $\rightarrow$ act coupling. The adjacent
transition probability is 0.002, and the unadjusted lift is 1.05, whereas the
adjusted OR is 1.33 [1.09, 1.62]. Actionability may still be desirable, but these
analyses provide no consistent behavioural evidence for the coupling, and our
observational design cannot show that improving actionability changes behaviour.

\textbf{Verifiability.} The claim that documentation should be written so agents
can verify their work against it describes no observed behaviour: zero validation
events. Verifiability therefore cannot be justified solely by appealing to the
behaviour observed in this corpus.

\textbf{Documentation as failure recovery.} The framing of documentation as the
primary resource to which agents turn when stuck is supported by neither the
trigger distribution (7.5\% failure-driven) nor the recovery analysis (5.4\% of
episodes).

\textbf{Ranking recovery strategies.} We explicitly decline to conclude that
documentation-based recovery is more effective, despite it having the highest
point estimate (63.6\%). With observable outcomes for only 11 episodes, the
interval spans 35.4--84.8\% and overlaps every alternative.

\subsection{For dataset and tool builders}

Two measurement observations generalise beyond this study. First, agents that
route file operations through shell commands hide their file access inside
command strings. Any corpus analysis based only on tool names will
systematically undercount such agents, and cross-agent comparisons will reflect
extraction coverage rather than behaviour. Second, agent-facing
documentation is invisible to file-type taxonomies built before 2024. Any
documentation classifier without \texttt{agent\_instruction} and
\texttt{agent\_working\_note} categories will place the majority of agent
documentation interactions in a residual bucket, as ours did before we added
them.

%===============================================================================
\section{Threats to Validity}
\label{sec:threats}

\subsection{Construct validity}

\textbf{Documentation is identified by file path.} Docstrings, inline comments,
and prose embedded in source files are invisible to our instrument. This
systematically undercounts documentation work, and the undercount is
non-uniform: languages and projects that favour in-source documentation are
underrepresented. Our absolute rates are therefore lower bounds. The comparative
findings --- agent-facing versus project documentation, read versus write ---
are affected only if in-source documentation is distributed very differently
across those categories than path-identified documentation, which we cannot rule
out.

\textbf{Purpose is not measured.} Our initial scheme included a purpose
dimension. We removed it because purpose is not recoverable from tool-call logs:
a file read is compatible with many intents, and assigning one would be
unfalsifiable. We report trigger, interaction type, and outcome instead. Thus, by
design, we do not analyse agents' reasons for reading particular documents.

\textbf{Tier-2 labels are unvalidated.} The \texttt{agent\_working\_note}
category --- 25.1\% of documentation events, and one of our headline findings ---
rests on language-model classification of 500 ambiguous paths (98.4\% of
ambiguous events), with 27 paths falling back to keyword rules. \emph{No human
validation of these labels has been performed.} The necessary next step is dual
human coding of a 200--300-event subsample, with inter-rater reliability measured
using Cohen's $\kappa$ or Krippendorff's $\alpha$. Until then, the precise share
of this category should be treated as provisional. The qualitative
finding that agent-authored working documents constitute a large and previously
uncategorised class is more robust than its exact magnitude, since it is visible
in the raw paths.

\subsection{Internal validity}

\textbf{Stage assignment is a sticky heuristic.} Once a failure signal appears, a
session remains in the debugging stage until a test or build passes. This inflates
the debugging share. We advance only the negative claim (documentation is not
confined to orientation) and do not interpret the stage distribution as a precise
allocation.

\textbf{Outcome detection uses regular expressions over tool output.} Our
success/failure signal is missing when output contains no recognisable indicator,
and missing is not the same as failure. Resolution rates are computed only for the 662 of
2{,}034 episodes with an observable outcome. If episodes without observable
outcomes differ
systematically from observable ones, the resolution rates would be biased by a
selection effect we cannot quantify.

\textbf{Transition probabilities are first-order.} A near-zero adjacent
transition from documentation read to code edit does not preclude longer-range
influence. The three-event-horizon lift analysis is our mitigation; its
unadjusted estimate shows no association (lift 1.05), although the adjusted model
does (OR 1.33 [1.09, 1.62]). Influence at longer ranges, or influence mediated
through reasoning that we cannot observe, would not be detected by either
analysis.

\textbf{Trigger assignment uses a fixed lookback.} Triggers are assigned using a
four-event window. A documentation read prompted by an event outside that window
would most likely be misattributed to agent initiative, which is our largest
category and therefore the one most exposed to this error.

\subsection{Extraction fidelity}

We validated extraction against the dataset's independently computed
\texttt{tool\_call\_count}; the counts matched exactly in five of six
spot-checked sessions. The extractor handles four transcript formats. We
identified and fixed
two defects during the study: failure to parse shell-embedded paths, which caused
one agent family to appear to have no documentation events, and failure to handle
non-string tool output, which excluded nine sessions. Both defects were detected
through implausible results rather than dedicated tests. Residual undercounting
likely remains for shell-centric agents, so we report per-agent rates as lower
bounds of varying tightness rather than as directly comparable estimates.

\subsection{External validity}

\textbf{SWE-chat is opt-in telemetry.} The sessions were contributed by
developers using agent CLIs, and 87\% of the corpus comes from a single agent family.
Developers who opt into sharing may be systematically more experienced, more
open-source oriented, or working on more shareable tasks than the population of
agent users.

\textbf{AIDev comprises public repositories that adopted agents early.} It
over-represents projects receptive to agentic contribution.

\textbf{Neither dataset necessarily generalises to private codebases.}
Documentation practices, review norms, and agent configurations may all differ
in private settings.

\textbf{The two datasets do not represent the same population.} We use them as
complementary evidence about processes and artefacts, respectively; we never pool
their units, and agreement between them corroborates a pattern but does not
constitute cross-validation of a measurement.

\textbf{The corpus is a snapshot of a fast-moving practice.} Agent-facing
documentation conventions are approximately two years old and changing. The
specific 60.5\% share characterises this snapshot, not a stable
constant; the finding we expect to persist is the existence and prominence of the
category, not its exact magnitude.

\subsection{Statistical validity}

Primary uncertainty estimates are cluster-bootstrap intervals with 2{,}000
resamples --- sessions resampled for SWE-chat statistics, repositories for AIDev
--- computed by the percentile method at fixed seeds, with each resample
recomputing the pooled proportion from summed within-cluster counts. Wilson
intervals accompany them as independence-assuming references only. Transition and
lift intervals use the same session-level procedure. We apply no multiple-comparison
correction across the strata examined (agent, language, task type, star bucket,
outcome), so small differences between adjacent strata should not be
overinterpreted; we base no claim on such a difference. Where a cell is small, we
state this limitation and avoid drawing a conclusion, as in the recovery-strategy ranking
($n=11$).

%===============================================================================
\section{Conclusion}
\label{sec:conclusion}

We measured how coding agents interact with technical documentation across 557 real
agentic sessions and 33{,}097 agentic pull requests. The dominant finding is
that agentic development has produced a genre that documentation research has
not yet studied: agent instruction files and agent working notes,
which together account for 60.5\% of observed documentation interaction and are
written by agents nearly as often as they are read. The documentation types that
current tooling and quality frameworks target --- API references and
troubleshooting documentation --- account for 1.3\% and 0.4\% of interactions,
respectively.

Three assumed mechanisms are not consistently supported by the data. The association between
consultation and subsequent code editing remains unresolved: unadjusted
estimates show no elevation, whereas stage-adjusted estimates do. Documentation
is rarely the first failure-recovery
resource: it is the first recovery move in 5.4\% of 2{,}034 failure episodes.
Finally, no explicit documentation-based validation sequence was observed, which means
verifiability --- a property routinely described as desirable for agent-facing
documentation --- corresponds to no behaviour recorded by our instrument and
must therefore be designed for rather than assumed.

The practical implication is a reallocation of effort: improving agent
instruction files has the potential to reach agents far more often than
equivalent work on API reference prose. The research implication is that
documentation now has two audiences with measurably different behaviours, and
the newer audience writes almost as much as it reads.

\section*{Data and Materials Availability}

Both datasets are public. SWE-chat is gated but freely available subject to acceptance
of its ODC-BY terms; we used the snapshot of 19 August 2026, comprising 5{,}851
sessions. Of these, 5{,}790 carry an agent label from one of the six families
with at least 20 sessions and form the base for agent reweighting; the remaining
61 comprise 52 sessions with no agent label and 9 from smaller families. AIDev is
openly downloadable; we used its curated subset.

The replication package contains the four format-specific extractors and the
20-symbol event alphabet; the Tier-1 path rules and Tier-2 labels for all 527
ambiguous paths; the event-level table of 3{,}033 coded interactions with
evidence spans; the cluster-bootstrap and GEE scripts with fixed seeds; and the
figure- and table-generation code, which generates every table directly from the
analysis outputs so no tabulated value is transcribed by hand.

\bibliographystyle{ACM-Reference-Format}
\bibliography{references}

\end{document}